\documentclass[prl,reprint, nofootinbib,amsmath,amssymb, aps, floatfix]{revtex4-1}
\usepackage{dcolumn}

\usepackage{bm}
\usepackage{amssymb,amsmath,amsthm,graphicx,ulem}
\usepackage[linktoc=page]{hyperref}

\usepackage{graphicx,subfigure}
\usepackage{epsfig}
\usepackage{amsmath}
\usepackage{amsfonts}
\usepackage{amssymb}
\usepackage[usenames]{color}
\usepackage{enumerate}

\usepackage[letterpaper,left=2.cm,right=2.cm,top=2.5cm,bottom=2.5cm]{geometry}
\usepackage[T1]{fontenc}

\theoremstyle{remark}

\numberwithin{equation}{section}
\begin{document}
\normalem

\title{A New Area Law in General Relativity}
\author{Raphael Bousso}%
\affiliation{Department of Physics, University of California, Berkeley, CA 94720, USA}
\affiliation{Lawrence Berkeley National Laboratory, Berkeley, CA 94720, USA}
\email{bousso@lbl.gov}
\author{Netta Engelhardt}
\affiliation{Department of Physics, University of California, Santa Barbara, CA 93106, USA}
\email{engeln@physics.ucsb.edu}
\bibliographystyle{utcaps}

\begin{abstract}
We report a new area law in General Relativity. A future holographic screen is a hypersurface foliated by marginally trapped surfaces. We show that their area increases monotonically along the foliation. Future holographic screens can easily be found in collapsing stars and near a big crunch. Past holographic screens exist in any expanding universe and obey a similar theorem, yielding the first rigorous area law in big bang cosmology. Unlike event horizons, these objects can be identified at finite time and without reference to an asymptotic boundary. The Bousso bound is not used, but it naturally suggests a thermodynamic interpretation of our result.
\end{abstract}

\maketitle




A black hole is a region from which no signal can escape to arbitrarily distant regions. Accordingly its event horizon is defined as a connected component of the boundary of the past of future infinity. This definition has proven fruitful, giving rise to laws of black hole ``mechanics''~\cite{Haw71,BarCar73} analogous to the laws of thermodynamics. We now recognize this analogy as a physical property~\cite{Bek72,Haw75}: black holes have entropy, and they radiate at a finite temperature.

\noindent {\bf The Teleological Problem \ } Note that the object behaving thermally is characterized only by the event horizon. By definition, its location is ``teleological'': it depends on the entire future history of the spacetime. This leads to some puzzles. Consider an observer surrounded by a massive shell collapsing at the speed of light. He sees only a perfectly flat spacetime region, before, during, and after crossing the event horizon. By causality, the thermodynamic properties of the event horizon must not have any physical manifestation in this region. Conversely, the formation of a black hole can be causally determined at finite time. At this point its thermodynamic properties, too, should be verifiable without knowledge of the future. Whether a collapsed star radiates today cannot depend on whether the black hole is greatly enlarged tomorrow. 

Similar remarks apply to Rindler or de Sitter horizons: they have thermodynamic properties~\cite{Unr76,GibHaw77a}, but their location is defined in terms of the distant future.

Thus it would be of great interest to identify a geometric object that is well-defined quasi-locally, and which obeys some or all of the laws of thermodynamics. By quasi-local, we mean that the object may have finite size but its definition only depends on its immediate spacetime neighborhood. Such an object would not only resolve the above puzzle. It could also extend the geometric thermodynamics to cosmological spacetimes, which need not have an asymptotic boundary.

\begin{figure}[ht]
\subfigure[]{
\includegraphics[width=0.23 \textwidth]{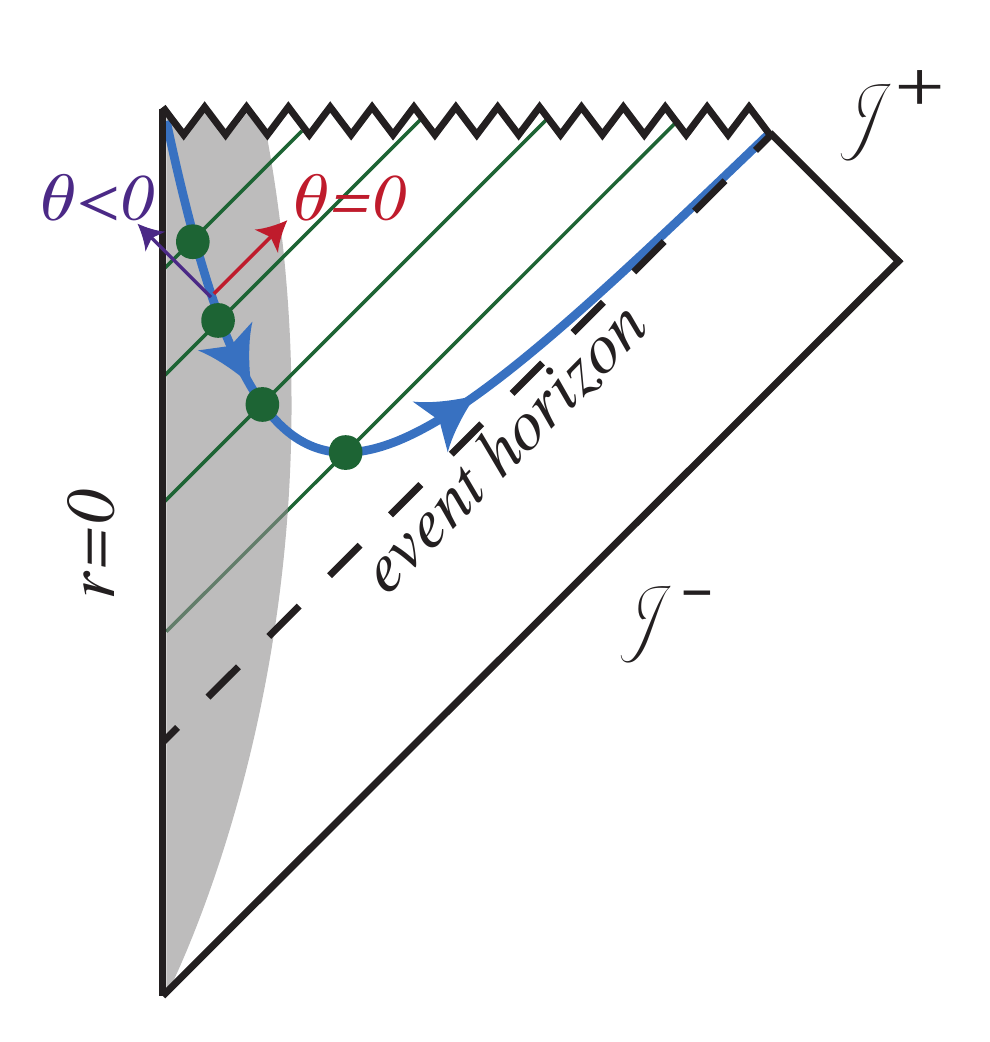}
}
\subfigure[]{
\includegraphics[width=0.22 \textwidth]{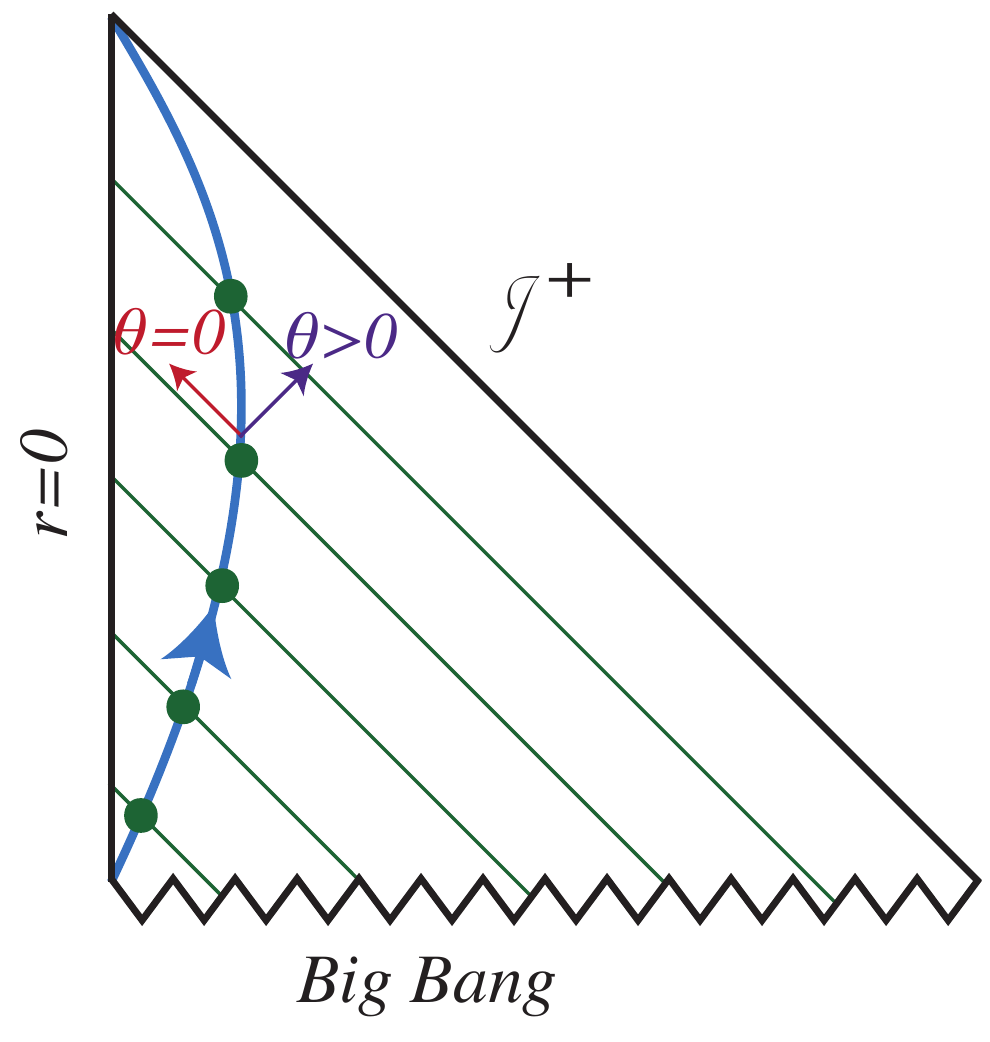}
}
\caption{To construct a past or future holographic screen (blue solid line), one chooses a null foliation of the spacetime (green diagonal lines) and finds the maximum area surface $\sigma$ on each null slice (green dots)~\cite{CEB2}. We prove that the area of the {\em leaves} $\sigma$ grows monotonically (in the direction of the blue triangle). (a) Black hole formed by collapse of a dust ball (grey shaded), with a null foliation by future light-cones of a worldline at $r=0$. Marginally trapped spheres (dots) form a {\em future holographic screen}. This is not a dynamical horizon: the screen is timelike inside the ball and becomes spacelike only where low-density dust falls in at late times (white region). (b) Expanding matter-dominated FRW universe, with a foliation by past light-cones of a comoving worldline. Marginally antitrapped spheres form a {\em past holographic screen}, which is everywhere timelike.}
\label{fig-examples}
\end{figure}
\noindent {\bf Future Holographic Screens \ } Extant proposals fall short~\cite{Booth05}. The term {\em apparent horizon} has been used in two inequivalent ways. With one definition it depends strongly on a choice of Cauchy foliation and does not obey an area theorem. With another it reduces to the event horizon in a broad class of examples~\cite{BenDov:2006vw} (and perhaps in all cases~\cite{Eardley:1997hk}) and thus shares its dependence on the future. A {\em Killing horizon} does not exist in general spacetimes and at best supplies an approximate notion in nearly stationary settings. 

A more promising structure is captured by the elegant notions of {\em future outer trapped horizon} (FOTH)~\cite{Hay93} and {\em dynamical horizon}~\cite{AshKri02}, which are equivalent in a classical, physical regime. A dynamical horizon $H_{\rm dyn}$ is a spacelike hypersurface foliated by marginally trapped surfaces $\sigma$ called {\em leaves}. A marginally (anti-)trapped surface is a codimension 2 compact spatial surface with vanishing expansion in one orthogonal future lightlike direction, ``outside,'' and strictly negative (positive) expansion in the other null direction, ``inside.'' The definition of $H_{\rm dyn}$ immediately implies that the area of its leaves grows in the outward direction.

However, the requirement that $H_{\rm dyn}$ be spacelike constitutes a crippling limitation. Consider the simplest semi-realistic black hole solution: the Oppenheimer-Snyder collapse of a pressureless homogeneous dust ball. Marginally trapped round spheres form a timelike hypersurface (not a dynamical horizon) in the collapsing star; see Fig.~\ref{fig-examples}a. They form a spacelike hypersurface (a dynamical horizon) only where the density of infalling matter is small (and in itself insufficient to create a black hole). This behavior is insensitive to small perturbations, and it arises in a wide class of collapse solutions~\cite{BoothBrits}. Timelike sections are also ubiquitous in cosmology, see Fig.~\ref{fig-examples}b. Thus, the notion of a dynamical horizon (or of a FOTH) appears to be inapplicable in perfectly reasonable physical settings. 

Timelike hypersurfaces foliated by marginally trapped surfaces (``timelike membranes'') also have monotonic area~\cite{Hay93,AshKri02}. These objects were not widely considered, perhaps because their area {\em decreases} towards the future, or because their signature seems to render them unsuitable as ``black hole boundaries''. In any case, as Fig.~\ref{fig-examples}a shows, one generically finds that marginally trapped surfaces form a hypersurface with mixed signature. The area growth or decrease on individual portions with definite signature has no clear significance---unless the pieces conspire to yield a single unified area law. This possibility has been remarked upon~\cite{BoothBrits}, but to our knowledge it was not pursued further.

Here we report a general area theorem for an entire hypersurface $H$, of indefinite signature. We require only that $H$ be foliated by marginally trapped  (or by marginally anti-trapped) surfaces $\sigma$. 
We will call $H$ a {\em future (or past) holographic screen}, in reference to the holographic properties conjectured by the Bousso bound~\cite{CEB1}, which indicates that such surfaces may have a thermodynamic interpretation. Such hypersurfaces have been referred to as {\em marginally trapped (or anti-trapped) tubes} in the dynamical horizon literature. However, the holographic interpretation was not recognized, nor was an area theorem proven. The significance of our construction is twofold: first, in proving that future holographic screens obey an area law; and second, in using the Bousso bound to propose a thermodynamic interpretation of this area law.

\noindent {\bf Theorem \ } {\em Let $H$ be a future holographic screen. Then the area of its leaves $\sigma$ increases strictly monotonically along $H$, under one continuous choice of flow.}

With the opposite choice, the area would decrease; the nontrivial point is that the evolution is monotonic. The direction of increase is easily seen to be the past (on timelike portions of $H$) or exterior (on spacelike portions). Thus, a key intermediate result established in the full proof is that $H$ evolves {\em only\/} into the past and/or exterior of each leaf. (It may flow to the past near some portions of a leaf and to the exterior near others.)

\noindent {\bf Sketch of Proof \ } We begin by considering the spacelike and timelike portions of $H$ separately.  Any spacelike portion of a future holographic screen is a dynamical horizon and hence has increasing area~\cite{AshKri02,Hay93}, in the outward direction. To see this, note that the flow from leaf to leaf can be thought of as an infinitesimal deformation away from one leaf along the future null congruence with vanishing expansion, and then back onto $H$ along the future-contracting (and thus, past-expanding) congruence orthogonal the next leaf. At linear order, the first step leaves the area invariant and the second increases it. By a similar argument, any timelike portion of $H$ has strictly increasing area in the past direction. 

Thus our task is nontrivial only if $H$ contains both timelike and spacelike portions.  To prove the theorem, we must constrain how its spacelike and timelike portions can meet. Some signature changes are consistent with monotonicity. For example, the area increases along the L-shaped future holographic screen in Fig.~\ref{fig-examples}a, which is initially timelike-past-directed and then becomes spacelike-outward-directed.  Other types of transitions would violate area monotonicity; these must be shown to be impossible. 
\begin{figure}[t]
\includegraphics[height=0.35 \textwidth]{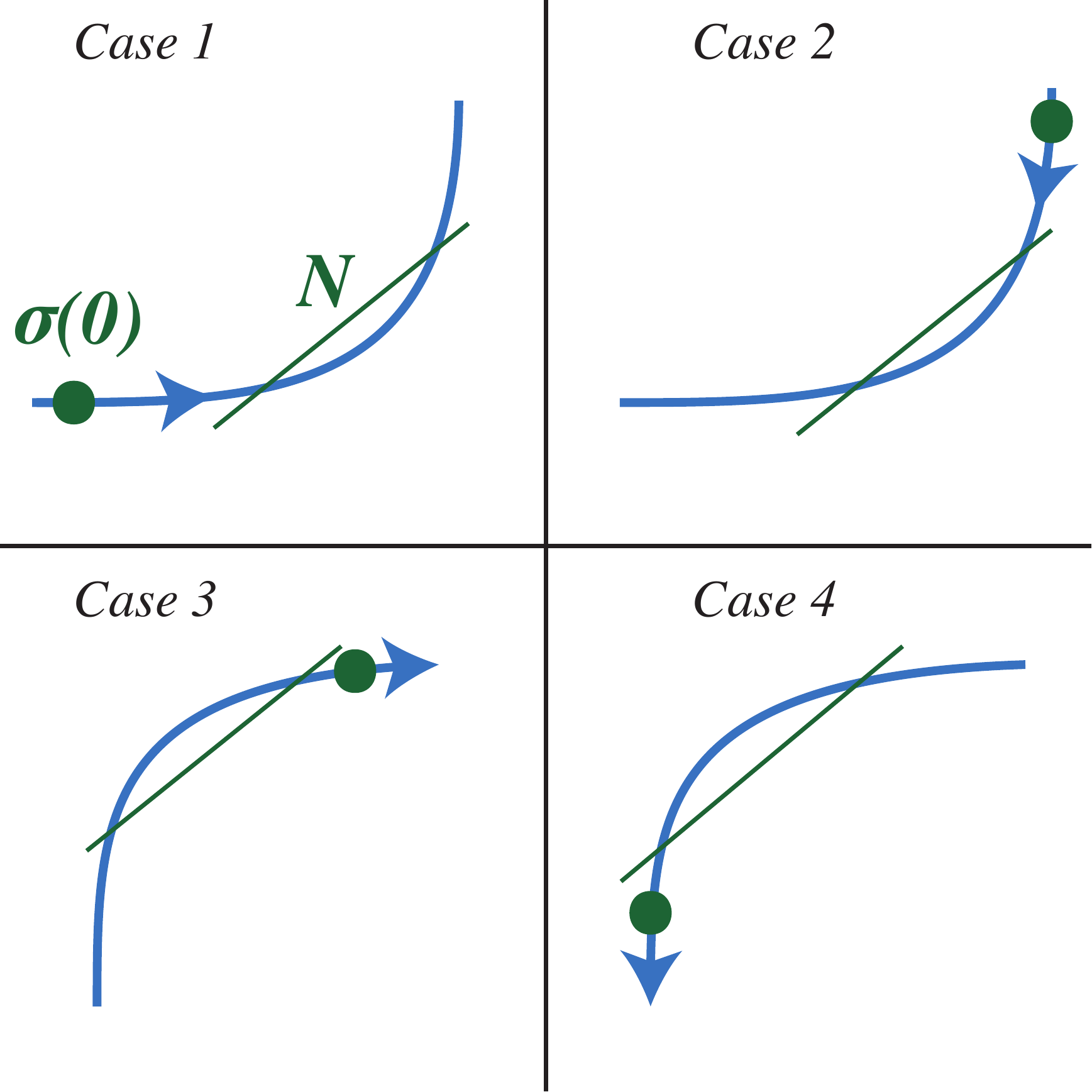}
\label{fig-sphsym-a}
\caption{There are four types of signature changes on a future holographic screen $H$ that would violate the monotonicity of the area. The dot indicates a reference leaf $\sigma(0)$, near which the area increases in the direction indicated by the arrow. On the far side of the ``bend'' it decreases. Note that $H$ is tangent to the marginal null direction ($\theta=0$) at the transition. Thus the marginal null hypersurface $N$ orthogonal to a nearby leaf intersects $H$ twice (diagonal line). In the classical, generic regime we consider, this is impossible.}
\label{fig-sphsym}
\end{figure}

Suppose we pick some leaf $\sigma(0)$ on $H$. Let us begin by following the leaves locally in the direction of increasing area, i.e., in the spacelike-outward or timelike-past direction. We must then show that further along the same direction of flow, the screen cannot ever become timelike-future-directed, or spacelike-inward-directed, since on such portions the area would decrease.  Similarly, flowing in the opposite direction from $\sigma(0)$, the area initially decreases under this reverse flow. We must show that no transitions occur that would cause it to increase eventually, along the reverse flow. Thus, there are four distinct types of monotonicity-violating transitions that must be ruled out; they are shown in Fig.~\ref{fig-sphsym}.

Our general proof is presented in a separate paper~\cite{BouEng15b}; it establishes several nontrivial intermediate results. The proof assumes the null curvature condition, $R_{ab} k^a k^b\geq 0$, where $R_{ab}$ is the Ricci tensor and $k^a$ is any null vector. This is appropriate for a classical regime. It implies that light-rays focus, i.e., the expansion $\theta$ at regular points of any null congruence is nonincreasing~\cite{Wald}. We also add some unrestrictive technical assumptions to the definition of $H$, in particular a generic assumption that excludes certain degenerate cases with no matter and shear. 

We offer here a simplified and less rigorous argument. It applies only to the special case of spherically symmetric spacetimes, but it captures important elements of the full proof. Consider a leaf $\sigma$ on a future holographic screen $H$ whose foliation respects the spherical symmetry. By the generic condition, the expansion $\theta$ immediately becomes negative in the two opposite marginal ($\theta=0$) null directions away from any leaf $\sigma$ of $H$. These two marginal congruences (future-outward and past-inward) generate two opposing light-sheets (null hypersurfaces with everywhere nonpositive expansion~\cite{CEB1}), $N^\pm(\sigma)$. Let $N(\sigma)$ be their union.

For contradiction, let us assume the existence of one of the transitions shown in Fig.~\ref{fig-sphsym} that would violate area monotonicity. Near the transition there exists a leaf $\sigma$ such that $N(\sigma)$ intersects $H$ not only at $\sigma$, but also on the far side of the transition from $\sigma$. This second intersection surface must again be a leaf of $H$, by spherical symmetry. But by the generic condition and the null curvature condition, it is impossible that $\theta=0$ on both leaves. This contradicts the defining property of $H$ that each leaf be marginally trapped.

It is easy to understand why the full proof must be more elaborate. Without spherical symmetry, the boundary between timelike and spacelike portions of $H$ will not generally coincide with a single leaf; nor will the intersection of an appropriate null surface $N$ with $H$ coincide with a leaf. This means that not all four types of monotonicity-violating transitions can be dealt with by a single argument: there remain two substantially different cases, each of which requires a more sophisticated argument than the one given above.

\noindent {\bf Relation to the Holographic Principle \ } Our generalization of dynamical horizons can be alternatively thought of as a refinement of the notion of {\em preferred holographic screen hypersurfaces}~\cite{CEB2}. These are foliated by marginal surfaces (i.e., one null expansion vanishes, but no sign is imposed on the other expansion). These objects arise naturally from the study of the covariant entropy bound (Bousso bound)~\cite{CEB1}, the conjecture that the area of any spacelike codimension 2 surface $B$, in Planck units, bounds the entropy of matter on any light-sheet orthogonal to $B$. A light-sheet is a null hypersurface with everywhere nonpositive expansion, $\theta\leq 0$, in the direction away from $B$. See~\cite{RMP} for a review.

There are four null congruences orthogonal to any surface $B$ (future and past-directed, ingoing and outgoing). In typical situations, two of the four have strictly negative expansion and so generate a light-sheet. (For a sphere in Minkowski space, this would be past and future inward directions; for a trapped sphere, e.g.\ inside a black hole, it would be the inward and outward future directions; and for an anti-trapped sphere, e.g.\ near the big bang, it would be the inward and outward past directions.) If the spacetime satisfies the null curvature condition, the light-sheet can be continued up to conjugate points (caustics) of the congruence. The Bousso bound constrains the entropy on each lightsheet. 

The remaining two null congruences must then have positive expansion, since they are the continuation of the same light-rays through $B$, in the opposite direction.  Their entropy would not be constrained by the area of $B$. However, if $B$ is marginal, i.e., if $B$ has one null direction with $\theta=0$, then we may consider the entire null hypersurface $N$ that contains $B$ and has vanishing expansion at $B$. $N$ can now be viewed as the disjoint union of two light-sheets. In this sense, the entropy on an entire null slice $N$, and not just on one ``one side'' of $B$, is constrained by the Bousso bound in the special case where $\theta=0$ on $B$.

It is easy to find a sequence of surfaces $B(r)$ so that the associated null hypersurfaces $N(r)$ foliate the spacetime. This is accomplished by a reverse construction~\cite{CEB2}: one begins by picking a null foliation, and on each slice $N$ one finds the spatial cross-section with maximum area (see Fig.~\ref{fig-examples}). The union of these maximal surfaces is the preferred holographic screen associated with the foliation.

The significance of this construction is as a precise statement of the sense in which all spacetimes are hologram. Loosely, the holographic principle~\cite{Tho93,Sus95} says that the number of degrees of freedom fundamentally scales like the area of surfaces, not the enclosed volume as one would expect from local field theory. It had been unclear how to make sense of this idea in cosmological spacetimes, particularly those without any boundary (e.g., a closed, recollapsing universe)~\cite{FisSus98}. The construction of preferred holographic screen hypersurfaces~\cite{CEB2} answered this question. It identified surfaces whose area, by the Bousso bound, limits the entropy of all matter in the spacetime, slice by slice. In some cases the screen hypersurface will lie on the conformal boundary of the spacetime, consistent with the locus of explicit holographic theories known for these settings~\cite{Mal97}. In cosmology and other regions where gravity is strong, the screen lies partially or wholly in the interior of the spacetime (Fig.~\ref{fig-examples})~\cite{CEB2,RMP}.

We have offered this discussion for context. An area law is of intrinsic interest, and we make no use of the above conjectures in its proof. However, it is significant that the theorem applies to surfaces whose area already admits an interpretation related to entropy. We intend to return to this connection in the near future.

\noindent {\bf Existence and Uniqueness \ }  Like dynamical horizons, holographic screens are highly nonunique: different null foliations yield different screens. Unlike dynamical horizons, they are known to exist broadly and are easily found (Fig.~\ref{fig-examples}). This was already known for the ``preferred holographic screen hypersurfaces'' of~\cite{CEB2}.  We have now added a refinement as described above, distinguishing between marginally trapped and anti-trapped leaves. This does not appear to diminish the abundance of holographic screens substantially. Future holographic screens are abundant inside black holes and in collapsing cosmologies.  Past holographic screens can be found for any big bang universe populated by matter, radiation, and vacuum energy, homogeneous or perturbed.

\noindent {\bf Discussion \ } Our result establishes the first broadly applicable area theorem in General Relativity since Hawking's 1971 proof that the future event horizon has nondecreasing area~\cite{Haw71}. Our theorem applies to all future or past holographic screens.

The admissability of timelike portions on holographic screens has an interesting consequence. If we interpret the area as an entropy, our area law implies that future holographic screens have a past-directed arrow of time. In this sense, at least, {\em time runs backwards inside a black hole and in collapsing universes}.  By contrast past holographic screens, which appear in expanding cosmologies, have a normal arrow of time on their timelike portions. Perhaps this puzzling result is related to recent arguments that sufficiently old black holes cannot have a smooth horizon~\cite{AMPS}.

Interpretations aside, any object that obeys a universal area theorem deserves our attention. It does so all the more in light of the conjectured significance of holographic screens in quantum gravity~\cite{CEB2}. Future and past holographic screens appear to give rise to a rich mathematical structure, whose investigation we initiate in our full proof~\cite{BouEng15b}. It will be important to understand the extent to which other laws of thermodynamics apply, particularly the first law. 

In a separate article we will describe the quantum extension of our results. The area of a cross-section of an event horizon admits a semi-classical generalization via the  substitution $A\to 4G\hbar S_{\rm gen}$~\cite{Bek72}. In fact, this can be applied to an arbitrary Cauchy-splitting two-surface $\sigma$, and thus the notion of (marginally) trapped surface can be given a quantum generalization as well~\cite{Wal10QST}. Here $S_{\rm gen}$ is the generalized entropy, i.e., the sum of the area of $\sigma$ and the matter entropy on one side of $\sigma$. 

As classical future and past holographic screens obey an area law, we expect that quantum holographic screens obey a Generalized Second Law. In particular, we expect the quantum extension of our results to yield a rigorous formulation of the Generalized Second Law in cosmology.\\

\noindent {\bf Acknowledgments \ } It is a pleasure to thank M.~Aganagic, D.~Engelhardt, S.~Fischetti, D.~Harlow, G.~Horowitz, W.~Kelly, S.~Leichen\-auer, T.~Jacobson, D.~Marolf, M.~Moosa, R.~Wald, and A.~Wall for discussions and correspondence. The work of RB is supported in part by the Berkeley Center for Theoretical Physics, by the National Science Foundation (award numbers 1214644 and 1316783), by fqxi grant RFP3-1323, and by the US Department of Energy under Contract DE-AC02-05CH11231. The work of NE is supported in part by the US NSF Graduate Research Fellowship under Grant No. DGE-1144085 and by NSF Grant No. PHY12-05500.

\bibliographystyle{utcaps}
\bibliography{all}
\end{document}